\documentclass[twoside]{revtex4}

\usepackage{amsfonts,amsmath,amssymb,bm,hyperref,graphicx}

\def\text{\rm}
\def\Brunt{Brunt-V\"{a}is\"{a}l\"{a}\ }

\begin{document}

%\preprint{E-print archive: astro-ph/}

\title{Coriolis force corrections to g-mode spectrum in 1D MHD model}

\author{Maxim Dvornikov$^{a,b}$}
\email{dvmaxim@cc.jyu.fi}
\author{Timur Rashba$^{a,c}$}
\email{timur@mppmu.mpg.de}
\author{Victor Semikoz$^{a}$}
\email{semikoz@yandex.ru}
\affiliation{$^a$IZMIRAN, 142190, Troitsk, Moscow region, Russia;\\
$^b$Department of Physics,
P.O.~Box~35, FIN-40014 University of Jyv\"askyl\"a, Finland;\\
$^c$Max-Planck-Institut f\"ur Physik, F\"ohringer Ring 6, D-80805, M\"unchen, Germany}

\date{\today}

\begin{abstract}
The corrections to g-mode frequencies caused by the presence of a central 
magnetic field and rotation of the Sun are calculated. The calculations are 
carried out in the simple one dimensional magnetohydrodynamical model using
the approximations which allow one to find the purely analytical spectra 
of magneto-gravity waves beyond the scope of the JWKB approximation and avoid 
in a small background magnetic field the appearance of the cusp resonance which
locks a wave within the radiative zone. These analytic results are compared with %%@
the 
satellite observations of the g-mode frequency shifts which are of the order one %%@
per 
cent as given in the GOLF experiment at the SoHO board. The main contribution turns 
out to be the magnetic frequency shift in the strong magnetic field which obeys the
used approximations. In particular, the fixed magnetic field strength 700 KG 
results in the mentioned value of the frequency shift for the g-mode of the 
radial order n=-10. The rotational shift due to the Coriolis force appears to be 
small and does not exceed a fracton of per cent, $\alpha_{\Omega}\leq 0.003$.       
\end{abstract}

\maketitle

%%%%%%%%%%%%%%%%%%%%%%%%%

\section{Introduction}

The recent progress in the search for low frequency solar oscillations --
possibly g-modes --
achieved in the GOLF (SoHO) experiment~\cite{TC04} and future 
experimental proposals~\cite{Turck-Chieze:2005im} open new possibilities for the
detailed studies of the solar radiative zone (RZ) as well as the solar core which
cannot be done by solving inverse helioseismic problem
for the high frequency p-modes.

The frequencies found in the experiment~\cite{TC04} turn out to be slightly shifted
$\delta \omega / \omega \sim 1\thinspace\%$ compared to the predictions of the
Solar Seismic Model (SSeM)~\cite{Turck-Chieze2001,Turck-Chieze2003} obtained in 
the absence of magnetic field 
and solar rotation.

The influence of the RZ magnetic field on the g-modes frequency shifts was
studied in Ref.~\cite{Rashba}. In that work the calculations were performed in the
simplified one dimensional (1-D) model that allowed one to evaluate the scope of
the magnetic field perturbation theory without g-modes absorption (and reflection)
by Alfven or cusp resonances layers inside the RZ. This simplification guarantees
that g-modes can enter the solar surface. Such 1-D calculations appeared to be
in a good agreement with the later magnetic shifts estimates in three dimensional
(3-D) MHD model~\cite{Rashba1}.

The correct separation of the cusp (Alfven) resonances in 1-D model (see
Sec.~\ref{Marcttdms} below), which enables one to simplify the problem in question, %%@
is
the significant advantage in comparison with 3-D model that allows us to compute %%@
rotation
and magnetic corrections. Such corrections are calculated in
a 3-D model with help of the perturbation theory, without referring to the presence
of MHD resonances, using eigenfunctions of horizontal and vertical displacements
$\xi_{h,r}^{(n,l)}$ computed in SSeM not accounting for a magnetic field or a
stellar rotation. On the contrary, the eigenfunctions of the 1-D problem, the %%@
vertical velocities
$v_z^{(n,l,\Omega,B_0)}=-i\omega(n,l,\Omega,B_0)\xi_z^{(n,l,\Omega,B_0)}(z)$,
depend on both the magnetic field and the stellar rotation. These functions are the
exact analytical solutions to the considered problem.

In the present work we continue to study the modifications of \emph{analytical} MHD
spectra for internal helioseismological waves in the 1-D model beyond the JWKB
approximation taking into account both the magnetic field and the \emph{rotation}
of the Sun. Note that in the 3-D MHD the solar rotation can be taken into 
account in analytical spectra only in the frame of the JWKB approximation 
of short waves, $L \ll c_s^2/g$
\cite{KumTalZah99}.

It should be also mentioned that in the 1-D model we loose the effect of the 
stellar rotation with respect to an observer because of the simplification of the
topology. In the 3-D model this effect is characterized by the azimuthal frequency
splitting $\sim m\Omega$. However, even in the case of the rigid body rotation, the
influence of the Coriolis force in the reference frame, rotating together with a
star, results in the additional frequency splitting which remains in the 1-D 
problem as well. The aim of this work is to evaluate such an effect in a
magneto-active plasma beyond the scope of the perturbation theory, applied in the
3-D model, or JWKB approximation. However we will use the exact analytical %%@
solutions
of the 1-D model.

This paper is organized as follows. In Sec.~\ref{BiMHDe} we formulate the MHD model %%@
for
an ideal plasma, discussing all the approximations used in Sec.~\ref{1Drma}. In
Sec.~\ref{Dr} we evaluate the importance of the differential rotation in RZ. Then
the approximation of the rigid body rotation is used in RZ. In Sec.~\ref{LiMHDirS} %%@
we
linearize the complete set of the MHD equations and derive the master equation for
the $z$-component of the velocity perturbation $v_z(z)$. With help of this velocity
component one can receive all the remaining MHD perturbed quantities including the
magnetic field perturbations. It is the component which results in the Doppler
shift of optical frequencies, measured in a helioseimic experiment. In
Sec.~\ref{Zmfl} we consider the case without the magnetic field, accounting for the
solar rotation only.

In Sec.~\ref{Marcttdms} we derive the simplified master equation for $v_z(z)$ in %%@
the limit
of relatively weak magnetic field that allowed one to avoid the MHD (cusp or
Alfven) resonances in the RZ of the Sun. Note that such a resonance was examined
Refs.~\cite{TC05,Burgess04a,Burgess03}. Using the magnetic field perturbation 
theory in Sec.~\ref{Marcttdms} we obtain the exact analytical g-modes spectrum in %%@
presence 
of magnetic fields and solar \emph{rotation} in RZ.

In Sec.~\ref{Discussion} we discuss our results.

\section{Basic ideal MHD equations}\label{BiMHDe}
We start from the following set of MHD equations. The mass
conservation law for the total density is of the form:
\begin{equation}\label{mass}
\frac{\partial \overline{\rho }}{\partial t}+\nabla \cdot \left( \overline{%
\rho }\mathbf{v}\right) =0, 
\end{equation}
where the total density is the sum of two terms, background density
and perturbation, $\overline{\rho }=\rho _{0}+\rho'$.

In the system rotating with the Sun (${\bf v}_0=0$) the momentum
conservation takes the form:
\begin{eqnarray}\label{Euler}
&&\frac{\partial {\bf v}}{\partial t} + ({\bf v}\nabla){\bf v} +
2(\bm{\Omega}\times {\bf v})=\nonumber\\&&
-\frac{1}{\bar{\rho}}\nabla \left(p + \frac{H^2}{8\pi}\right) +
\frac{1}{8\pi \bar{\rho}}({\bf H}\nabla){\bf H} +{\bf g} - {\bf
\Omega}\times (\bm{\Omega}\times {\bf r})+\nonumber\\&& + {\bf
r}\times \left[\frac{\partial \bm{\Omega}}{\partial t} + ({\bf
v}\nabla)\bm{\Omega}\right],
\end{eqnarray}
where the total magnetic field ${\bf H}={\bf B}_0 + {\bf B'}$ includes
the background magnetic field ${\bf B}_0$ and the magnetic field
perturbation ${\bf B'}$, ${\bf g}$ is the gravity acceleration, ${\bf
\Omega}$ is the angular rotation frequency.

In ideal MHD the Faraday equation takes the form
\begin{equation}\label{Faraday}
\frac{\partial {\bf H}}{\partial t}=({\bf H}\cdot \nabla ){\bf v}-({\bf v}%
\cdot \nabla ){\bf H}-{\bf H}u , 
\end{equation}
where $u=\nabla \cdot {\bf v}\neq 0$ is the compressibility of gas.

Finally the full nonlinear ideal MHD equation system is completed by
the equation of the entropy conservation:
\begin{equation}\label{entropy}
\frac{\partial }{\partial t}\left( \frac{\overline{p}}{\overline{\rho
} ^{\gamma }}\right) + ({\bf v}\cdot \nabla )\left( \frac{\bar{p}}{
\overline{\rho }^{\gamma }}\right)=0,
\end{equation}
where $\bar{p}=p_0 + p'$ is the total gas pressure; $\gamma=c_p/c_V$
is the heat capacity ratio ($\gamma=5/3$ for hydrogen plasma).  

Note, that in the equilibrium condition,
\begin{equation}\label{equilibrium}
-\frac{1}{\rho_0}\nabla \left(p_0 + \frac{B_0^2}{8\pi}\right) + {\bf
  g} - \bm{\Omega}\times (\bm{\Omega}\times {\bf r})=0,
\end{equation}
the contribution of the centrifugal force (last term) is negligible in
the Sun ($\sim 2\times 10^{-5}$).

\subsection{1D rectangular model assumptions}\label{1Drma}

In order to find spectra of seismic waves accounting for the magnetic field
and rotation in RZ a number of
assumptions is required. We shall specify these as they are used,
in order to keep clear which results rely on which assumptions,
but we also list them all here for ease of reference. 
\begin{enumerate}
\item We consider ideal MHD neglecting both the heat conductivity and
  viscosity contributions to energy losses, as well as the ohmic dissipation.
\item We linearize the MHD equations about an equilibrium background
  configuration, {\it i.e.} a background configuration which is time
  independent and given by the condition~(\ref{equilibrium}).
\item We assume the fluctuations to be adiabatic, with the contributions of
  fluctuations to the heat source vanishing: $Q'=0$.
\item Moreover, we consider the fully ionized ideal gas, that means the
  thermodynamic quantity, first adiabatic exponent $\gamma = c_p/c_V$, is time
  independent and uniform. For numerical estimates we will take $\gamma=5/3$
  for hydrogen plasma.
\item We adopt the Cowling approximation, which amounts to the neglect of
  perturbations of the gravitational potential, ({i.e.:} $\phi' = 0$).
\item We assume the rigid rotation in RZ.
\item We use a rectangular geometry with the Cartesian coordinates:
  $x$, $y$, and $z$, where $z$ corresponds to the solar radial
  direction. See Fig.~\ref{fig:geometry}. The background quantities
  vary along the $z$ direction only (which implies the local
  gravitational acceleration, ${\bf g}$, is directed along the $z$
  axis, but in opposite direction). We also take a constant uniform
  background magnetic field, ${\bf B}_0$, pointing along the
  $x$-axis. The rotation (the angular velocity vector $\bf \Omega$) is
  assumed to be directed along the $y$-axis that is perpendicular to
  eqliptic ($x0z$-plane).  Far enough from the solar center such a
  slicy 1D configuration corresponds to a large scale toroidal
  magnetic field in RZ in the 3D geometry.
\begin{figure}[h]
\includegraphics[scale=.45]{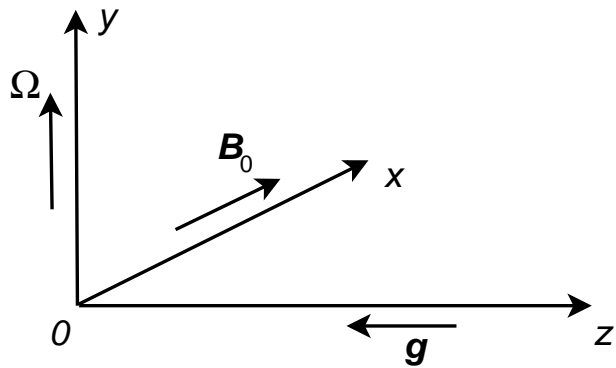}
\caption{\label{fig:geometry} The rectangular (1D) geometry.}
\end{figure}
\item The background mass-density profile is assumed to be exponential,
  $\rho_0 = \rho_c \, \exp[- z/H]$, for constant $\rho_c$ and $H$.  The
  conditions of hydrostatic equilibrium for the background then determine the
  profiles of thermodynamic quantities, and in particular imply $\gamma$ is a
  constant.
\item We assume that Brunt-Vaisala frequency is zero in CZ and non-zero, but
  constant, in RZ.
\end{enumerate}

This last assumption applies to very good approximation for
real mass-density profiles obtained by Standard Solar Models,
provided we identify the $z$ direction with the radial direction,
and focus our attention to deep within the radiative zone. The
constancy of $\gamma$ in this region is also expected since the
highly-ionized plasma satisfies an ideal gas equation of state to
good approximation. The rectangular geometry provides a reasonable
description so long as we do not examine too close to the solar
centre. What is important about our choice for ${\bf B}_0$ is that
it is slowly varying in the region of interest, and it is
perpendicular to both ${\bf g}$ and all background gradients,
$\nabla \rho_0$, $\nabla p_0$, {\it etc}.

As suggested in~\cite{Burgess04a} such 1D picture can be
fully described in analytical terms in contrast to the 3D case.

% There are two parameters which describe the spectra of magneto-gravity
% waves~\cite{Burgess04a}: (i) strength of the background magnetic field
% $B_0$ and (ii) the dimensionless transversal wave number $k=k_x
% H$. Here $H$ is the density scale height.

\subsection{Differential rotation}\label{Dr}
In 3D-geometry (spherical coordinates), for which $\bm{\Omega}=\Omega
\cos\theta {\bf e}_r -\Omega\sin \theta {\bf e}_{\theta}$, in the
absence of angular acceleration, $\partial \Omega/\partial t=0$, the
last term in Eq.~(\ref{Euler}) , $-r\sin \theta({\bf v\nabla})\Omega
{\bf e}_{\Phi}$, is much less than the Coriolis term in the l.h.s. of
Eq.~(\ref{Euler}) under the condition
$$ r\sin \theta\left(v_r\frac{{\partial \Omega}}{\partial r} +
\frac{v_{\theta}}{r}\frac{\partial \Omega}{\partial \theta}\right)\ll
2\Omega\left[\sin \theta v_r + \cos \theta v_{\theta}\right],
$$
or
\begin{equation}\label{differ3D}
r\frac{\partial \Omega}{\partial r}\ll 2\Omega,\quad \tan
\theta\frac{\partial \Omega}{\partial \theta}\ll 2\Omega.
\end{equation}
There are no doubts that such conditions can be fulfilled in RZ (for
almost rigid rotation) while in CZ the last inequality is valid near
equator $\theta= \pi/2$, (inclined on 7 degrees with respect to the
eqliptic latitude $\vartheta=0^{\circ}$), only in the limit $\partial
\Omega/\partial \theta|_{\theta=\pi/2}\to 0$. Looking at the solar
rotation profile in Fig.~1 (right panel) in Ref.~\cite{TC05}
demonstrating the SoHO results we see that the first condition in
(\ref{differ3D}) is valid at least for low and middle latitudes with
an accuracy $\sim \Delta \Omega/\Omega\sim 50/400=0.125\ll 1$.

In our rectangular geometry the corresponding conditions: 
\begin{equation}\label{differ1D}
z\frac{{\rm d}\Omega}{{\rm d}z}\ll 2\Omega,~~~x\frac{{\rm
d}\Omega}{{\rm d}z}\ll 2\Omega\frac{v_x}{v_z},
\end{equation} 
are similar to (\ref{differ3D}) where the velocity ratio is larger
than the unity, $v_x/v_z> 1$, especially at the MHD resonances where
$v_x\gg v_z$.

In what follows we put $\Omega= const$, or neglect angular
acceleration and the differential rotation omitting the last line in
Eq.~(\ref{Euler}).

\section{Linear ideal MHD in rotating Sun}\label{LiMHDirS}
Assuming the plane wave propagation along the background magnetic
field, $\sim e^{-i(\omega t - k_xx)}$, in linear ideal MHD we obtain
from the full MHD system (\ref{mass})- (\ref{entropy}) the following
set of eight (=8) equations:
\begin{equation}\label{mass1}
\rho'=\frac{v_z}{i\omega}\frac{{\rm d}\rho_0}{{\rm d}z} +
\frac{\rho_0}{i\omega}\left(\frac{\partial v_y}{\partial y} +
\frac{\partial v_z}{\partial z}\right) + \frac{\rho_0v_x}{V},
\end{equation}
\begin{equation}\label{Eulerx}
P=\rho_0 V v_x + \frac{B_0B_x'}{4\pi} - \frac{2\rho_0\Omega v_z}{ik_x},
\end{equation}
\begin{equation}\label{Eulery}
\frac{\partial P}{\partial y}=i\rho_0\omega v_y
+\frac{ik_xB_0}{4\pi}B_y',
\end{equation}
\begin{equation}\label{Eulerz}
\frac{\partial P}{\partial z}=i\rho_0\omega v_z
+\frac{ik_xB_0}{4\pi}B_z' + 2\rho_0\Omega v_x - \rho' g,
\end{equation}
\begin{equation}\label{Bx}
B_x'=\frac{B_0}{i\omega}\left(\frac{\partial v_y}{\partial y} +
\frac{\partial v_z}{\partial z}\right),
\end{equation}
\begin{equation}\label{By}
B_y'=-\frac{B_0}{V}v_y,
\end{equation}
\begin{equation}\label{Bz}
B_z'=-\frac{B_0}{V}v_z,
\end{equation}
\begin{equation}\label{entropy1}
\rho'=\frac{1}{c_s^2}\left[P - \frac{B_0B_x}{4\pi}\right] +
\frac{v_z}{i\omega}\frac{(\gamma -1)}{\gamma}\frac{{\rm d}\rho_0}{{\rm
d}z},
\end{equation}
where $V=\omega/k_x$ is the phase velocity of MHD oscillations we are
looking for; $P=p' + B_0B_x'/4\pi$ is the total pressure perturbation
in linear approximation; $c_s=\sqrt{\gamma gH}$ is the sound velocity;
$H=0.1R_{\odot}$ is the density scale height in the background density
$\rho_0=\rho_ce^{-z/H}$, $\rho_c=150~g/cm^3$ is the central mass
density in the Sun.

Using Eq. (\ref{Eulerx}), from which one finds $P-B_0B_x'/4\pi$, then
equating (\ref{mass1}) and (\ref{entropy1}) we derive $v_x$ in terms of
$v_{y,z}$-components:
\begin{equation}\label{vx}
v_x=\frac{c_s^2V}{V^2-c_s^2}\left[\frac{1}{i\omega}\left(\frac{\partial
v_y}{\partial y} + \frac{\partial v_z}{\partial z}\right) +
\frac{v_z}{i\omega \gamma \rho_0}\frac{{\rm d}\rho_0}{{\rm d}z} +
\frac{2\Omega v_z}{ik_xc_s^2}\right].
\end{equation}
We see from Eq. (\ref{Eulery}) that in our geometry we can put
$v_y=B_y'=\partial/\partial y=0$, or RZ medium is uniform along
rotation axis ($y$-axis). This allows us with the use of
Eq. (\ref{Eulerx}) for the pressure $P$, Eq. (\ref{Bx}) for $B_x'$ and
$v_x$ from Eq. (\ref{vx}) to obtain from Eq. (\ref{Eulerz}) the master
differential equation of the second order:
\begin{eqnarray}\label{master}
&&\zeta(1-\zeta)\frac{{\rm d}^2v_z}{{\rm d}\zeta^2} - \zeta\frac{{\rm
d}v_z}{{\rm d}\zeta} +\nonumber\\&&+ k^2\left[ 1 +
\zeta^{-1}\left(\frac{N^2}{\omega^2}-1 +\frac{\omega^2 -
4\Omega^2}{k_x^2c_s^2} + \frac{2H\Omega}{Vk^2}\left[\frac{2}{\gamma}
-1\right]\right)\right]v_z=0.\nonumber\\
\end{eqnarray}
Here we input the \Brunt frequency $N$, $N^2=(c_s^2/H^2)(\gamma
-1)/\gamma^2=(g/H)(\gamma -1)/\gamma$, on which we can normalize the
compressibility parameter
$a_1=\omega^2/k_x^2c_s^2=0.24(\omega/N)^2/K^2$ where the coefficent
$0.24=(\gamma -1)/\gamma^2$ is given by the heat capacity ratio
($\gamma=5/3$) for hydrogen plasma. The argument $\zeta=(1 - a_1)\xi$,
$\xi=\xi_0e^{z/H}$, is given by the background magnetic field $B_0$,
$\xi_0=k_x^2v_{A0}^2/\omega^2$, where $v_{A0}$ is the Alfv\'en
velocity at the center of the Sun, $v_{A0}=(B_0/43.4~G)~{\rm cm/sec}$,
given by the central density $\rho_c$.

The solution of this equation can be easily obtained from the Gauss
equation for $v_z=\zeta^{\sigma}Y(\zeta)$ in analogy with
\cite{Burgess04a}:
\begin{equation}\label{Gauss}
v_z(\zeta)=D_1\zeta^{\sigma_1}F(a,b;c;\zeta) + D_2\zeta^{1-\sigma_1 -
\gamma^{-1}}F(1+a-c,1+b-c;2-c;\zeta),
\end{equation}
where $F(a,b;c;\zeta)$ are Gauss (hypergeometric) functions,
\begin{equation}\label{roots}
\sigma_{1,2}=\frac{1}{2}\pm \frac{1}{2}\sqrt{1 +
4K^2\left[\frac{N^2}{\omega^2} -1 + \frac{\omega^2 -
4\Omega^2}{k_x^2c_s^2} + \frac{2\Omega}{\omega
k}\left(\frac{2}{\gamma} -1\right)\right]}
\end{equation}
are the upperscripts in the solution (upper sign for $\sigma_1$),
$a=\sigma_1 + K$, $b=\sigma_1 - K$, $c=a+b$, $K=k_x H$.

Using the resonable boundary conditions (see e.g. in
\cite{Burgess04a}) one can derive the MHD {\it dispersion equation}
for the magneto-gravity modes $\omega (K,B_0, \Omega)$ expressed
through hypergeometric functions in (\ref{Gauss}) at the fixed point
(at the center of the Sun, $z=0$,
$\zeta=\zeta_0=(1-a_1)\xi_0$). However, the analysis of such
dispersion equation is still very complicated.  Below we consider some
simplifications of the master equation~(\ref{master}) which allow us
to get analytic MHD spectra in the RZ.

\subsection{Zero magnetic field limit}\label{Zmfl}

\begin{figure}[h]
\includegraphics[scale=.45]{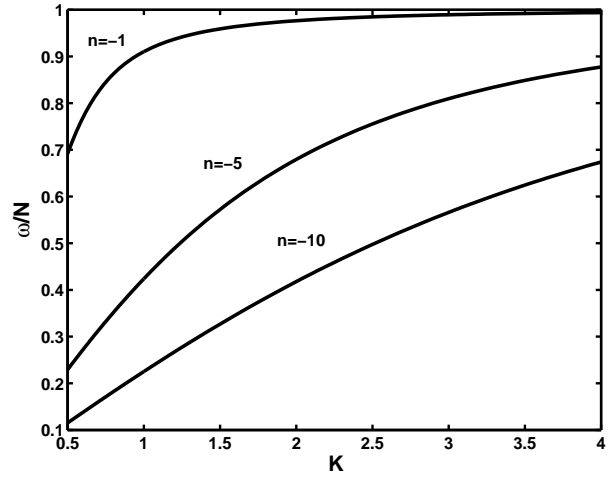}
\caption{The $g$-mode spectra accounting for the solar
  rotation versus wave number, $K$, shown for different mode numbers,
  $n=-1,-5,-10$.\label{fig:omegaN}}
\end{figure}

In the limit $\zeta=(1-a_1)\xi\to 0$, $B_0\to 0$, with the use of the
transformation $v_z=e^{z/2H}\Psi(z)$ and the change $\zeta{\rm d}/{\rm
d}\zeta=H{\rm d}/{\rm d}z$, Eq.~(\ref{master}) converts to:
\begin{equation}\label{Bzero}
\frac{{\rm d}^2\Psi}{{\rm d}z^2}
+\left[\frac{\omega^2-4\Omega^2}{c_s^2}- \frac{1}{4H^2}
+k_x^2\left(\frac{N^2}{\omega^2} -1 + \left(\frac{2}{\gamma}
-1\right)\frac{2\Omega}{\omega k_xH} \right)\right]\Psi=0.
\end{equation}
In the absence of rotation, $\Omega=0$, this equation is similar to
3D osillation Eq.~(7.90) in~\cite{Christensen03}:
\begin{equation}\label{3D}
\frac{{\rm d}^2\Psi}{{\rm d}r^2} +\left[\frac{\omega^2}{c_s^2}-
\frac{1}{4H^2} +k_h^2\left(\frac{N^2(r)}{\omega^2}
-1\right)\right]\Psi=0.
\end{equation}
Rewriting Eq. (\ref{Bzero}) for $\Omega=0$ as
$$ 
\frac{{\rm d}^2\Psi}{{\rm d}z^2} + \frac{\beta_0^2}{4H^2}\Psi(z)=0,
$$
where $\beta_0=\sqrt{4K^2(N^2/\omega^2 -1) - 1 + 4K^2a_1}$ in RZ, and as
$$
\frac{{\rm d}^2\Psi}{{\rm d}z^2} - \frac{\Gamma_0^2}{4H^2}\Psi(z)=0,
$$
in CZ, where $N_{CZ}=0$, or $\beta_0\to i\Gamma_0=i\sqrt{4K^2(1-a_1)
+1}$, then matching the corresponding solutions at the top of RZ,
$z=z_{RZ}$, one can easily obtain the g-mode spectrum in our 1D
model~\cite{Rashba}:
\begin{equation}\label{gmode}
\frac{\beta_0z_{RZ}}{2H} + \pi n=~{\rm
arccot}\left[-\frac{\Gamma_0}{\beta_0}\right].
\end{equation}
Here $n=-1,-2,-3,...$.  
The analogous solution of Eq. (\ref{Bzero})
for the rotating Sun, $\Omega\neq 0$, leads to the spectrum:
\begin{equation}\label{rotationspectrum}
\frac{\beta_{\Omega}z_{RZ}}{2H} + \pi n=~{\rm
arccot}\left[-\frac{\Gamma_{\Omega}}{\beta_{\Omega}}\right],
\end{equation}
where
\begin{equation}\label{betaOmega}
\beta_{\Omega}=\sqrt{0.96\left[\left(\frac{\omega}{N}\right)^2-\left(\frac{2\Omega}%%@
{N}\right)^2\right]-
1 +4k^2\left[\frac{N^2}{\omega^2} -1 +
\frac{1}{5k}\left(\frac{2\Omega}{N}\right)\left(\frac{N}{\omega}\right)\right]},
\end{equation}
\begin{equation}\label{GammaOmega}
\Gamma_{\Omega}=\sqrt{4k^2\left[1 - \frac{\omega^2
-4\Omega^2}{k_x^2c_s^2} -
\frac{1}{5K}\left(\frac{2\Omega}{\omega}\right)\right] +1}.
\end{equation}
Here $n=-1, -2,,...$ (see Fig.~\ref{fig:omegaN}).  We can neglect a
term in $\Omega^2$ in Eqs.~(\ref{betaOmega}), (\ref{GammaOmega}),
since the rotation parameter $2\Omega/N=2\times 10^{-3}$ is small
comparing with the frequencies in the GOLF experiment~\cite{TC04},
$1>\omega/N\geq 0.2$, we are interesting in.

The rough estimate of the rotation frequency splitting $\omega\to
\omega - \Delta \omega$ due to the Coriolis force near $\omega\leq N$,
can be easily found from Eq. (\ref{betaOmega}), accounting for the
shift of the parameter $N^2/\omega^2\to (N^2/\omega^2)(1 + 2\Delta
\omega/\omega)$:
\begin{equation}\label{shift}
\left|\alpha_{\Omega}\right|\equiv \frac{\Delta \omega}{\omega}=
\frac{1}{10K}\left(\frac{2\Omega}{N}\right)\left(\frac{\omega}{N}\right)=
\frac{1}{5K}\times
10^{-3}\left(\frac{\omega}{N}\right).
\end{equation}
Thus, for $K>1$ and waves $v_z$ propagating in eqliptic (where ${\bf
g}\perp \bm{\Omega}$) this frequency shift is negligible,
$\alpha_{\Omega}\sim 10^{-4}$.

\begin{figure}[h]
\includegraphics[scale=.45]{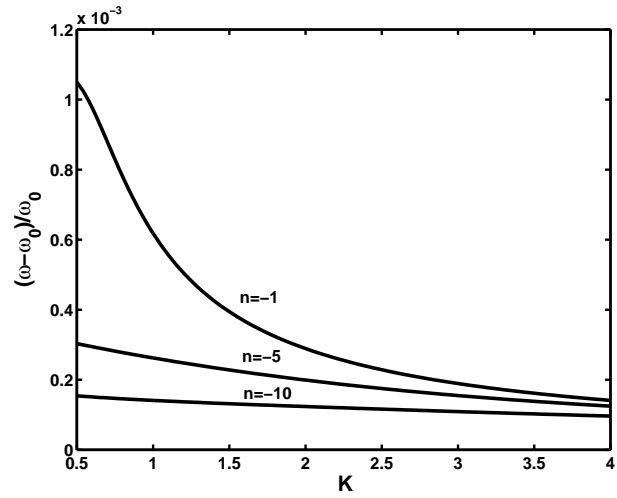}
\caption{\label{fig:alphaomega} The relative shift of g-mode frequency due
  to rotation, $(\omega-\omega_0)/\omega_0$, versus wave number, $K$,
  shown for different mode numbers, $n=-1,-5,-10$.}
\end{figure}

In Fig.~\ref{fig:omegaN} we show the g-mode spectrum $\omega(n,K)$
accounting for rotation $\Omega$ in the Sun and given by the
dispersion equation (\ref{rotationspectrum}) for a few modes,
$n=-1,-5, -10$.  In Fig~\ref{fig:alphaomega} we show $g$-mode
frequency shifts $\alpha_{\Omega}(K)=(\omega - \omega_0)/\omega_0$ in
dependence of the transversal wave number $K$ neglecting RZ magnetic
field, ${\bf B}_0=0$. Here the frequency $\omega_0$ responds to the
absence of rotation, $\Omega=B_0=0$. One can see that the rotation
correction originated by the Coriolis force increases if $K$
decreases, especially for the highest frequency $n=-1$. This behaviour
is in an agreement with what one expects in the 3D case (see
discussion below).


\subsection{Magnetic and rotation corrections to the g-mode %%@
spectrum}\label{Marcttdms}  
In order to obtain spectra of g-modes in the presence of the magnetic
field let us define the coefficent in front of the second derivative
in Eq.~(\ref{master}) as $1-\zeta$, where $\zeta=v_A^2(1-a_1)/V^2$.

We consider the perturbative regime for magnetic fields, where
 Alfv\'en velocity is much less than phase velocitiy,
 $v_A^2=v_{A0}^2e^{z/H}\ll
 V^2=\omega^2/k_x^2=[N^2H^2/K^2](\omega/N)^2$, so that $\zeta\ll 1$
 and MHD resonances ($\zeta=1$) do not appear within the RZ (see
 Fig.~2 in~\cite{Rashba}). The perturbative magnetic field for which
 the maximum Alfv\'en velocity is small, $v_A^2(0.7R_{\odot})\ll V^2$,
 obeys
$$
B_0\ll \frac{28}{K}\left(\frac{\omega}{N}\right)~{\rm MG}.
$$ 
This condition constrains the range of magnetic field strength when
the perturbative approach is valid. One can check that this condition
is well satisfied for the magnetic fields and for the modes down to
$n=-10$ considered here.

The absence of MHD resonances allows us to set
$$ v_z=\left[1 - \frac{v_A^2}{V^2}\left(1 -
\frac{V^2}{c_s^2}\right)\right]^{-1/2}e^{z/2H}\Psi (z),
$$
and thus derive from the master Eq.~(\ref{master}) the
$O(v_A^2/V^2)$-correction to Eq.~(\ref{Bzero}):
\begin{equation}\label{corr}
\frac{{\rm d}^2\Psi}{{\rm dz}^2} +\left[\frac{\beta_{\Omega}^2}{4H^2}
+ k_B^2(\Omega)\right]\Psi=0,
\end{equation}
where we defined $\beta_{\Omega}$ in Eq.~(\ref{betaOmega}) and the
wave number correction $k_B^2(\Omega)$ is given by
\begin{equation}\label{kB}
k_B^2(\Omega)=k_x^2(1-a_1)\frac{v_A^2}{V^2}\left[\frac{N^2}{\omega^2}
+ a_1 -\left(\frac{2\Omega}{N}\right)^2\left(\frac{0.24}{K^2}\right)+
\frac{1}{5K}\left(\frac{2\Omega}{N}\right)\left(\frac{N}{\omega}\right)\right].
\end{equation}
Introducing the coefficent
\begin{equation}\label{A0}
A_0=(1-a_1)\frac{v_{A0}^2}{V^2}\left[\frac{N^2}{\omega^2} + a_1
  -\left(\frac{2\Omega}{N}\right)^2\left(\frac{0.24}{K^2}\right)+
  \frac{1}{5K}\left(\frac{2\Omega}{N}\right)\left(\frac{N}{\omega}\right)\right],
\end{equation} 
and using the change of the variable $2s=z/H + \ln (4K^2A_0)$, one can
rewrite Eq.~(\ref{corr}) as
\begin{equation}\label{corr2}
\frac{{\rm d}^2\Psi}{{\rm ds}^2} +\left[\beta_{\Omega}^2 + e^{2s}\right]\Psi(s)=0.
\end{equation}
This equation has general solution expressed in terms of Bessel functions
of the first kind, $\Psi (s)=C_1J_{i\beta_{\Omega}}(e^s) +
C_2J_{-i\beta_{\Omega}}(e^s)$ where $C_{1,2}$ are constants. Then
following \cite{Rashba} we can find the corresponding solution in RZ
using the boundary condition at the solar center, $\Psi(z=0)=0$
(as $v_z(0)=0$), excluding one constant to obtain the solution of
equation (\ref{corr2}) in RZ, $0\leq z\leq z_{RZ}$:
\begin{eqnarray}\label{RZsolution}
&&\Psi_{RZ}(z)=C_{RZ}\Bigl[J_{i\beta_{\Omega}}(2K A_0^{1/2}e^{z/2H})
  -\nonumber\\&&- \frac{J_{i\beta_{\Omega}}(2K
    A_0^{1/2})}{J_{-i\beta_{\Omega}}(2K
    A_0^{1/2})}J_{-i\beta_{\Omega}}(2K A_0^{1/2}e^{z/2H})\Bigr]
\end{eqnarray}

In CZ the \Brunt frequency $N$ vanishes, $N=0$, so that we are led to
the same solution as in isotropic case, $\Psi_{CZ}(z)=C_{CZ}\exp (-
z\Gamma_{\Omega}/H)$, where $\Gamma_{\Omega}$ is given by
Eq.~(\ref{GammaOmega}), and we safely neglected the possible influence
of a CZ magnetic field (see comment on that after Eq.~(25)
in~\cite{Rashba}).  Since the arguments of Bessel functions are small,
we can use the first terms in the Bessel function series only.

Then by matching the logarithmic derivatives of the solutions
$\Psi_{RZ}$ and $\Psi_{CZ}$ at the top of RZ, $z=z_{RZ}$, one obtains
the dispersion equation for the case $B_0\neq 0$ which generalizes our
result in Eq. (27) \cite{Rashba} accounting for {\it rigid rotation}
of RZ in the Sun:
\begin{eqnarray}\label{main}
&&\beta_{\Omega}\left(1 + \frac{2\kappa^2_{RZ}}{1 +
\beta_{\Omega}^2}\right)\left[1 -
\beta_{\Omega}\frac{2\kappa^2_{RZ}}{1 + \beta_{\Omega}^2}\frac{1}{\sin
(\beta_{\Omega}z_{RZ}/H)}\right]\times\nonumber\\&& \times \cot
\left[\beta_{\Omega}\frac{z_{RZ}}{H}\right] - \frac{2\kappa^2_{RZ}}{1
+ \beta_{\Omega}^2}=-\Gamma_{\Omega},
\end{eqnarray}
where $\kappa_{RZ}=K A_0^{1/2}e^{z_{RZ}/2H}$, coefficents $A_0$,
$\beta_{\Omega}$ and $\Gamma_{\Omega}$ are given by Eqs.~(\ref{A0}),
(\ref{betaOmega}) and~(\ref{GammaOmega}) respectively.

\begin{figure}[h]
\includegraphics[scale=.45]{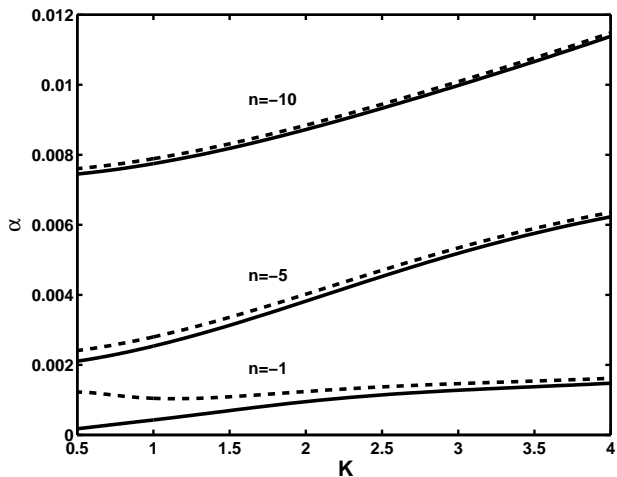}
\caption{\label{fig:figure_3} The relative shift of g-mode frequency due
  to rotation and magnetic field, versus wave number, $k$, shown for
  magnetic field $B_0=700$~kG and different mode numbers,
  $n=-1,-5,-10$. Solid lines -- neglecting rotation, dashed lines --
  accounting for the rotation.}
\end{figure}

This is our main equation giving the g-mode spectrum $\omega_B(K)$ in
1D MHD model accounting for the rotation of the Sun. In the
Fig.~\ref{fig:figure_3} we illustrate the rotation splitting
originated by the Coriolis force for the magnetic field $B_0=700~{\rm
kG}$ in dependence on the transversal wave number $K$ for a few modes
$n=-1, -5, -10$ (dashed lines there).

\section{Discussion}\label{Discussion}

As we expected the contribution of the Coriolis force to the g-mode
frequency shift is small: even for the radial order $n=-1$ one obtains
$\alpha_{\Omega}< 0.003$. In Fig.~\ref{fig:figure_3} we show how the
magnetic field shift $\alpha_B$ \cite{Rashba} is corrected due to the
presence of the solar rotation, $\Omega\neq 0$. We compare there two curves
calculated for the magnetic field $B_0=700~{\rm kG}$: (i)
$\alpha_B(K)=(\omega_{B0} - \omega_0)/\omega_0$ given by the solid
line and corresponding to the previous result obtained in
\cite{Rashba} in the absence of rotation, $\Omega=0$, and
(ii) $\alpha_{\Omega}(K)=(\omega_B - \omega_0)/\omega_0$ given by the dashed
line where $\omega_B$ accounts for the magnetic field and the Coriolis
force contributions.

Note that in our 1D approach, in which azimuthal number
$m$ is absent and the transversal wave number $K$ is analogous to the
angular degree $l$ (the Legendre number), the decrease of this wave number leads
to an enhancement of the Coriolis force contribution, as it can be seen on %%@
Fig.~\ref{fig:figure_3}, in the rotational splitting as it is predicted in the 3D %%@
case.

Let us note also that a crude approximation for the \Brunt frequency profile
$N(z)$ in RZ relies here on a fixed value $N=\text{constant}$ for which we have %%@
chosen the value $N=10^{-3}~\text{rad}/\text{sec}$ as in Ref.~\cite{Rashba}. 
This value was used in all calculations instead of
the maximum value of the \Brunt frequency, $N=2.8\times
10^{-3}~\text{rad}/\text{sec}$ ($\nu_N= N/2\pi\simeq 440~\mu \text{Hz}$ in cycling
frequencies, see, e.g., in Ref.~\cite{Christensen03}). Relying on such
maximum $N$ one gets $2.8^2 =7.8$ times smaller effect for the same
value of magnetic field. And, vice versa, the same magnetic shift of the order
1 \% (together with the Coriolis force correction)  can be obtained relying on the %%@
\Brunt frequency, $N=2.8\times 10^{-3}~\text{rad}/\text{sec}$, if a stronger (2.8 %%@
times greater) magnetic field exists in RZ. For example, to receive the %%@
corresponding frequency shift for the g-mode with $n= -10$ the magnetic field %%@
should be $B_0=700~{\rm kG}$.
 
Therefore magnetic fields resulting in the shift $\delta\omega/\omega \sim 1\%$ %%@
appear to be quite strong, in particular for high frequency g-modes, close to the %%@
\Brunt frequency, observed in the experiment. Thus for the corresponding shift of %%@
the $g_3^2$ mode with $\nu=222.46\thinspace\mu\text{Hz}$, one predicts the strong %%@
RZ magnetic field $B\geq 8\thinspace\text{MG}$ (see Ref.~\cite{Rashba1}) that %%@
already exceeds the existing central magnetic field limits. As it follows from the %%@
solar poles oblateness $\sim 10^{-5}$, one gets the limit $B\leq %%@
8\thinspace\text{MG}$ (see Ref.~\cite{FriGru04}). This discrepancy can be explained %%@
by the fact that the observed $g_3^2$ mode shift results from its interaction with %%@
other g-modes, close to the \Brunt frequency, rather than a magnetic field or a %%@
rotation.   

The configurations with 3-vectors ${\bf g}$, $\bm{\Omega}$ and ${\bf B}_0$, %%@
different from that shown on Fig.~\ref{fig:geometry}, in
the same 1-D rectangular geometry could lead to the different spectra we do not %%@
discuss here.  

In conclusion we mention that the ultimate goal of all solar studies would be the
complete 3D MHD description of the solar evolution. However, it is
rather far future from now. On the way to this goal in this letter we
present qualitatve analytical 1D model which describes the influence
of RZ magnetic field on helioseismic wave spectra accounting for the
solar rotation.

\begin{acknowledgments}
MD has been supported by the Academy of Finland under the 
contract No.~108875 and by a grant of the Russian Science Support Foundation. 
TR has been supported by the Marie Curie International Incoming Fellowship of 
the European Community. VS is very thankful S. Turck-Chi\'eze and A. %%@
Brun for helpful discussions. 
\end{acknowledgments}

%%%%%%%%%%%%%%%%%%%%%%%%%%%%%%%%%%%%%%%%%%%%

\end{document}